\documentclass[lettersize,journal]{IEEEtran}
\usepackage{amsmath,amsfonts}
\pdfoutput=1

\usepackage{cite}
\usepackage{graphicx}
\usepackage{textcomp}
\usepackage{xcolor}

\usepackage{algorithm}
\usepackage{algorithmic}
\usepackage{tcolorbox}

\usepackage{listings, xcolor}

\definecolor{verylightgray}{rgb}{.97,.97,.97}

\lstdefinelanguage{Solidity}{
	keywords=[1]{anonymous, assembly, assert, break, call, callcode, case, catch, class, constant, continue, constructor, contract, debugger, default, delegatecall, delete, do, else, emit, event, experimental, export, external, false, finally, for, function, gas, if, implements, import, in, indexed, instanceof, interface, internal, is, length, library, log0, log1, log2, log3, log4, memory, modifier, new, payable, pragma, private, protected, public, pure, push, require, return, returns, revert, selfdestruct, send, solidity, storage, struct, suicide, super, switch, then, this, throw, transfer, true, try, typeof, using, value, view, while, with, addmod, ecrecover, keccak256, mulmod, ripemd160, sha256, sha3}, 
	keywordstyle=[1]\color{blue}\bfseries,
	keywords=[2]{address, bool, byte, bytes, bytes1, bytes2, bytes3, bytes4, bytes5, bytes6, bytes7, bytes8, bytes9, bytes10, bytes11, bytes12, bytes13, bytes14, bytes15, bytes16, bytes17, bytes18, bytes19, bytes20, bytes21, bytes22, bytes23, bytes24, bytes25, bytes26, bytes27, bytes28, bytes29, bytes30, bytes31, bytes32, enum, int, int8, int16, int24, int32, int40, int48, int56, int64, int72, int80, int88, int96, int104, int112, int120, int128, int136, int144, int152, int160, int168, int176, int184, int192, int200, int208, int216, int224, int232, int240, int248, int256, mapping, string, uint, uint8, uint16, uint24, uint32, uint40, uint48, uint56, uint64, uint72, uint80, uint88, uint96, uint104, uint112, uint120, uint128, uint136, uint144, uint152, uint160, uint168, uint176, uint184, uint192, uint200, uint208, uint216, uint224, uint232, uint240, uint248, uint256, var, void, ether, finney, szabo, wei, days, hours, minutes, seconds, weeks, years},	
	keywordstyle=[2]\color{teal}\bfseries,
	keywords=[3]{block, blockhash, coinbase, difficulty, gaslimit, number, timestamp, msg, data, gas, sender, sig, value, now, tx, gasprice, origin},	
	keywordstyle=[3]\color{violet}\bfseries,
	identifierstyle=\color{black},
	sensitive=true,
	comment=[l]{//},
	morecomment=[s]{/*}{*/},
	commentstyle=\color{gray}\ttfamily,
	stringstyle=\color{red}\ttfamily,
	morestring=[b]',
	morestring=[b]"
}

\usepackage{graphicx}
\usepackage{subcaption}
\usepackage{url}
\usepackage{hyperref}
\usepackage{booktabs}
\usepackage{multirow}
\graphicspath{{picture}}
\usepackage{enumitem}
\usepackage{dashrule}
\usepackage{amsmath}
\usepackage{pifont}

\begin{document}

\title{Defining and Detecting the Defects of the Large Language Model-based Autonomous Agents}

\author{Kaiwen Ning, Jiachi Chen, Jingwen Zhang, Wei Li, Zexu Wang, Yuming Feng, Weizhe Zhang, Zibin Zheng,~\IEEEmembership{Fellow,~IEEE}
\IEEEcompsocitemizethanks{
            \IEEEcompsocthanksitem Kaiwen Ning, Jingwen Zhang, Zexu Wang are with the School of Software Engineering, Sun Yat-sen University and Peng Cheng Laboratory, China.
            \protect\\
            E-mail: \{ningkw, zhangjw273, wangzx97\}@mail2.sysu.edu.cn    \IEEEcompsocthanksitem Jiachi Chen, Zibin Zheng and Wei Li are with the School of Software Engineering, Sun Yat-sen University, China. \protect\\
			E-mail: \{chenjch86@mail, zhzibin@mail,liwei378@mail2\}.sysu.edu.cn
            \IEEEcompsocthanksitem Weizhe Zhang is with the School of Computer Science and Technology, Harbin Institute of Technology, China, and Peng Cheng Laboratory, China.\protect\\
			E-mail: wzzhang@hit.edu.cn 
            \IEEEcompsocthanksitem Yuming Feng is with the Department of New Networks, Peng Cheng Laboratory, China.\protect\\
            E-mail: fengym@pcl.ac.cn }
		}

\maketitle

\begin{abstract}

Artificial intelligence (AI) agents are systems capable of perceiving their environment, autonomously planning and executing tasks. Recent advancements in Large Language Models (LLMs) have introduced a transformative paradigm for AI agents, enabling them to interact with external resources and tools through prompt techniques. This advancement has significantly extended the capabilities of LLMs, positioning LLM-based AI Agents as an important research area. In such agents, the workflow integrates developer-written code, which manages framework construction and logic control, with LLM-generated natural language that enhances dynamic decision-making and interaction. 
However, discrepancies between developer-implemented logic and the dynamically generated content of LLMs in terms of behavior and expected outcomes can lead to defects, such as tool invocation failures and task execution errors.
These issues introduce specific risks, leading to various defects in LLM-based AI Agents, including service interruptions and incorrect output. 
Despite the importance of these issues, there is a lack of systematic work that focuses on analyzing LLM-based AI Agents to uncover defects in their code.
To address this gap, we present the first study focused on identifying and detecting defects in LLM Agents. We collected and analyzed $6,854$ relevant posts from StackOverflow to define and classify eight types of agent defects. For each defect type, we provided detailed descriptions and illustrated with an example. 
Then, we designed a static analysis tool, named Agentable, to detect the defined defects. 
Agentable leverages Code Property Graphs (CPGs) and LLMs to analyze Agent workflows by efficiently identifying specific code patterns and analyzing natural language descriptions.
To evaluate Agentable, we constructed two datasets: AgentSet, which consists of 84 real-world Agent projects, and AgentTest, which contains 78 Agent projects specifically designed to include various types of defects.
Our results show that Agentable achieved an overall accuracy of $88.79\%$ and a recall rate of $91.03\%$. Furthermore, our analysis reveals the $889$ defects of the Agent projects in the real-world dataset, highlighting the prevalence of these defects.

\end{abstract}

\section{Introduction}
\label{sec:intro}

Large Language Models (LLMs), with their powerful capabilities in text generation and semantic understanding, have introduced a new paradigm for artificial intelligence agents~\cite{agentsurvey1,agentsurvey4}. Unlike standalone LLMs, LLM-based Agents leverage prompt engineering techniques such as ReAct~\cite{yao2022react} and Chain of Thought (CoT)~\cite{turpin2024cot} to enable LLMs to perceive and utilize external resources and tools~\cite{agentsurvey2}. This not only mitigates the inherent knowledge lag in LLMs but also extends their functionality~\cite{agentsurvey3}. According to recent studies, LLM-based Agents have become one of the most popular research areas~\cite{agentsurvey5,agentsurvey6}. Additionally, an increasing number of LLM agent products and development frameworks have been put into production~\cite{agentsurvey7,chen2024agentpoisonredteamingllmagents}. For example, Langchain~\cite{guan2024langchain} and LLamaIndex~\cite{pedro2024llamaindex}.

However, since the workflow of Agents is controlled by both developer-written code and real-time natural language output from LLMs, the coupling between the various components of an Agent tends to be loose and abstract~\cite{li2024personal}. This can lead to unexpected behaviors during runtime, e.g., service interruptions~\cite{tang2024prioritizing}. Many developers in open-source communities have expressed concerns about this issue. For instance, an AI engineer from the well-known AI company Octomind reported numerous problems with Agents during both code development and actual execution~\cite{langchaindiss}. 

Unfortunately, there is still a lack of knowledge and tools to mitigate these threats. A recent survey~\cite{tang2024prioritizing} pointed out that the unique nature of the agent workflow can introduce various potential risks and called for researchers to pay attention to the possible security issues. Although some current studies have examined the reliability of Agents, they primarily focus on the applicability and security concerns, such as operational efficiency~\cite{liu2024a}, ethical standards~\cite{zhang2024s}, and backdoor attacks~\cite{wang-etal-2024-badagent}. However, few studies have addressed the code defects within Agent workflows. Code defects in an Agent refer to design flaws or issues within the Agent system, such as insufficient integration between the LLM and external tools. This can lead to the failure of the LLM to properly invoke external tools, causing the Agent to malfunction. As a result, characterizing Agent code defects and developing effective mitigation strategies remain open challenges.

To bridge this gap, we conducted the first study focused on analyzing design defects in LLM Agents and how to detect them. To define and categorize common Agent code defects, we collected $6,854$ relevant posts from StackOverflow and performed an empirical analysis to investigate the Agent code defects reported in these posts. Using a hybrid card sorting method~\cite{conrad2019making,laoshiCordSort}, we introduced the first systematic classification scheme for LLM Agent defects. This classification includes eight types of defects, covering external tools~\cite{mei2024llm}, LLM invocations~\cite{agentsurvey8}, memory mechanisms~\cite{yu2024finmem}, and planning modules~\cite{zhu2024knowagent}.

Based on our defect definitions, we propose Agentable, an LLM-powered static analysis approach for detecting potential code defects in real-world Agent projects. To the best of our knowledge, this is the first security technique targeting specific defects in Agent source code. It combines Code Property Graphs (CPGs)~\cite{wang2023graphspd} with LLMs to address the challenges posed by the complex development patterns of Agents in real-world environments. Specifically, Agentable first constructs a unified node (class, function) relationship tree from the CPG to improve the speed and accuracy of locating specific code, such as Tool initialization classes and their execution functions. Then, Agentable builds a semantic enrichment module based on the source code and its AST to capture complex semantic information related to Agent code defects, such as fault-tolerance mechanisms in specific code contexts. Finally, Agentable constructs an LLM generalization module to handle tasks requiring generalized recognition or natural language analysis, such as assessing the consistency between a Tool’s description and its name. With this design, Agentable effectively addresses the challenges posed by the complex and evolving development patterns of real-world Agents while maintaining strong generalization capabilities.

To evaluate the effectiveness of Agentable, we constructed AgentSet, a dataset consisting of $84$ real-world Agent projects. Additionally, to assess the recall rate of Agentable, we manually created AgentTest, a dataset of $78$ Agent projects, each labeled with specific defects, based on a high-quality open-source Agent project~\cite{basellm}. The experimental results show that Agentable achieved an overall accuracy of $88.79\%$ and a recall rate of $91.03\%$. Furthermore, based on Agentable’s detection results, we analyzed the prevalence and distribution of defects in real-world Agent projects.

We summarize our main contributions as follows:
\begin{itemize}
    \item We conducted the first study on code defects in LLM Agents. By analyzing $6,854$ relevant posts on StackOverflow, we defined and classified eight types of Agent code defects, expanding the current body of research on LLM Agent security~\cite{tang2024prioritizing,zhang2024agentsecurity,debenedetti2024agentsecurity}. We provided detailed descriptions of these defects, their triggering patterns, and potential solutions.
    \item We proposed Agentable, the first tool designed to detect code defects in Agents. This tool extracts fine-grained code semantics using Code Property Graphs (CPGs) and enhances generalization checks and natural language processing capabilities with LLMs, addressing the challenges posed by complex Agent code patterns in real-world environments.
    \item We constructed AgentSet, a manually curated dataset of $84$ real-world Agent projects. Additionally, we created AgentTest, the first dataset specifically designed for Agent code defects, consisting of $78$ Agent instances labeled with distinct defects. We evaluated Agentable on both datasets, and the results demonstrate its effectiveness, achieving an overall accuracy of $88.79\%$ and a recall rate of $91.03\%$.
    \item We will release the source code of Agentable and the related datasets after the paper is accepted to support further research.
\end{itemize}

 \section{Background and Motivation}
\label{sec:background}

\subsection{LLM-based Agent}
\label{sec2:Agent}

\begin{figure}[h]
	\centering{\includegraphics[scale=0.43]{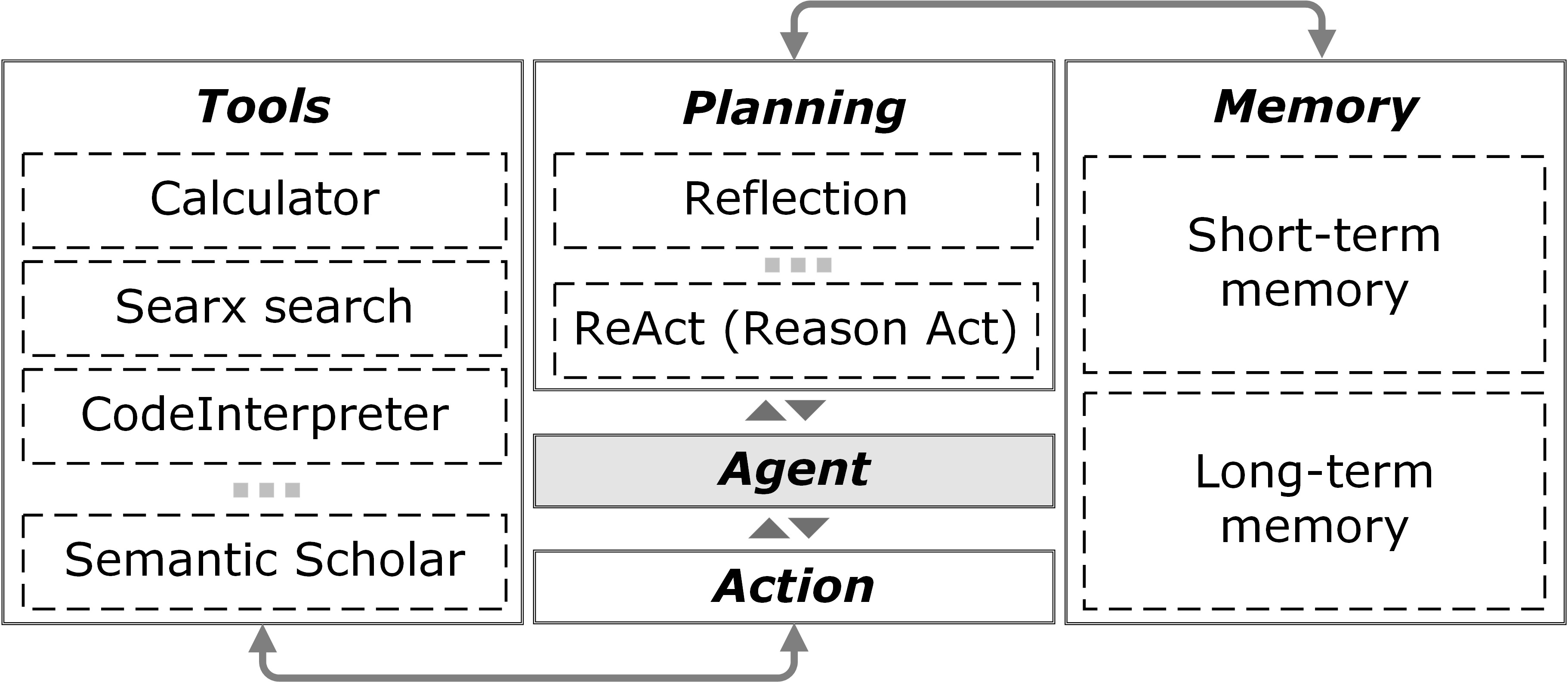}}
	\caption{An example of LLM-based Agent.}
	\label{Fig:agentexample}
\end{figure}

According to previous studies, a standard Agent typically consists of four components, i.e., \textit{planning, tools, memory}, and \textit{action}~\cite{agentsurvey6}. The \textit{planning} component, controlled by the LLM, is the core of the Agent’s decision-making process, interacting with \textit{tools} and \textit{memory} through the \textit{action} component to achieve specific goals. As shown in Fig.~\ref{Fig:agentexample}, OpenAI provides a model of an Agent~\cite{baseagent}. A simple Agent workflow can be summarized as follows: developers use prompt engineering techniques, such as Chain of Thought (CoT) or ReAct, to guide the LLM in formulating a plan~\cite{agentsurvey2}. While the LLM creates the plan, the Agent perceives the available list of tools and memory, and during the action phase, it invokes external tools or accesses the memory library as per the plan. Based on the results from the tools or memory, the Agent continues to call the LLM to refine the plan~\cite{agentsurvey5}.

In this simple Agent workflow, every action taken by the LLM is controlled by both the natural language output of the LLM and the developer-written code~\cite{baseagent}. This can lead to insufficient integration between the components of the Agent~\cite{chen2024agentpoisonredteamingllmagents}(see the example in \ref{Sec2:Mot}). Even more concerning, the Agent must interact with complex environments at runtime, such as invoking external tools, which may introduce additional unexpected situations. Therefore, it is crucial to analyze and inspect the Agent’s workflow code to safeguard against potential errors.

\subsection{Code Property Graphs}
\label{Sec2:CPG}

Graph-based code analysis methods typically transform source code into various graph structures, such as Abstract Syntax Trees (ASTs), Control Flow Graphs (CFGs), and Program Dependency Graphs (PDGs), to efficiently capture local structures and dependencies within the code~\cite{lee2023adcpg}. However, each type of graph focuses on different aspects, and no single graph can fully represent the semantics and logic of a program. The Code Property Graph (CPG) unifies multiple concepts, such as ASTs, CFGs or Evaluation Order Graphs (EOGs), Data Flow Graphs (DFGs), and Control Dependency Graphs (CDGs), into a single supergraph~\cite{yamaguchi2014modeling}. The advantage of this approach is that the CPG contains more relevant information for static program analysis while also providing a certain level of abstraction over the corresponding programming language.

\subsection{Motivation Example}
\label{Sec2:Mot}
\begin{figure}[h]
	\centering{\includegraphics[scale=0.2]{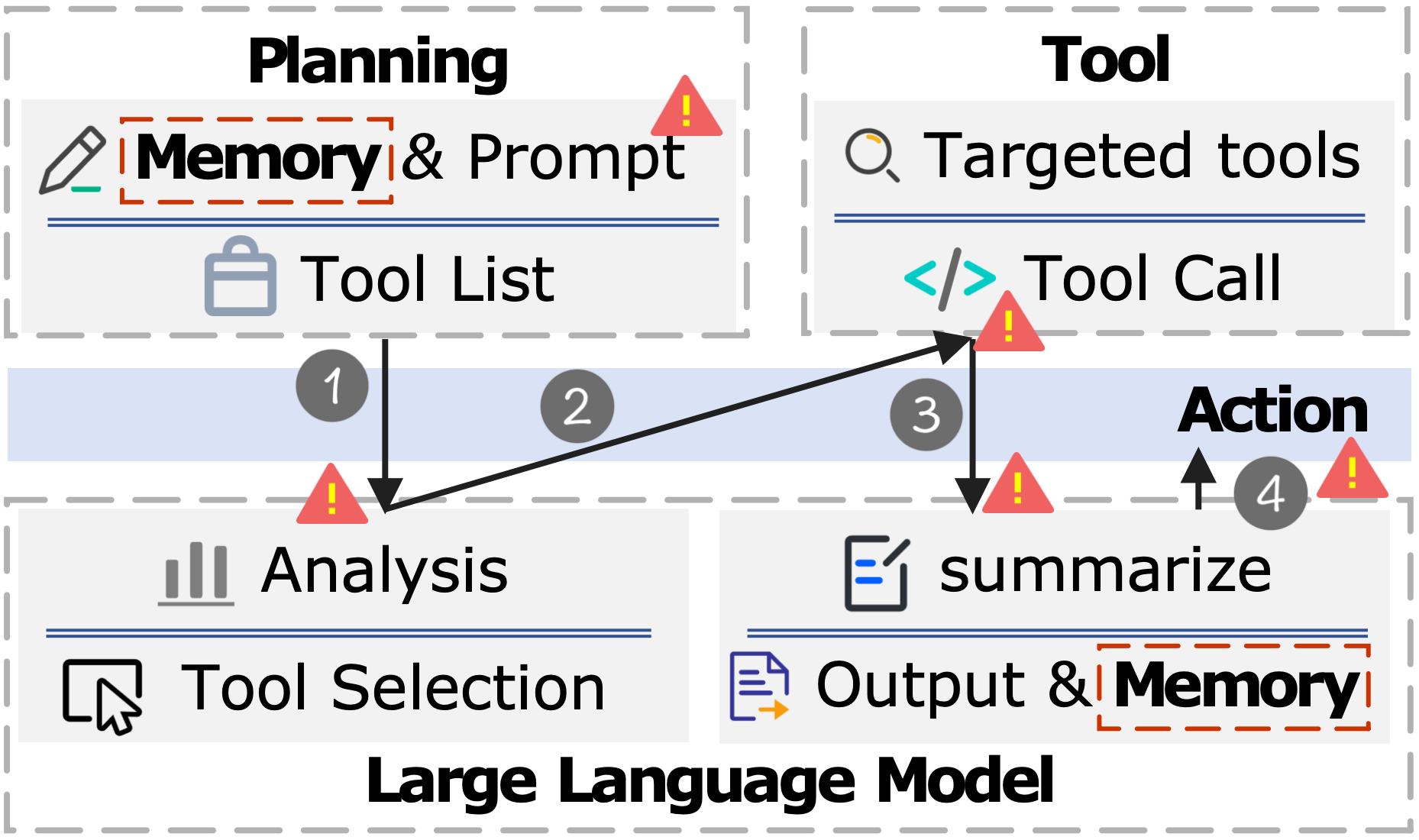}}
	\caption{An example of the motivation.}
	\label{Fig:motexample}
\end{figure}

In this part, we use an example to illustrate the complexity of the LLM-based Agent workflow and to emphasize the significance of this work. For example, as shown in Fig.~\ref{Fig:motexample}. This is an example of an Agent workflow where tasks are completed by invoking external tools. In the first step, the Agent’s Planning component assembles the user prompt, memory, and toolset into an input, which is passed to the LLM component through the Action component. Then, the LLM component analyzes the incoming information and selects the appropriate tool. In the second step, the LLM generates a decision and a list of parameters for invoking the tool, which is passed to the Tool component through the Action component. It is important to note that the output from the LLM is in natural language. In the third step, the Tool component receives the invocation information, selects the appropriate tool from the toolset, executes the corresponding tool call, and returns the result to the LLM via the Action component. In the fourth step, the LLM reflects on the results obtained from the Tool and returns the final decision to the Action component.

Throughout this process, the Agent relies on the natural language output from the LLM to select the appropriate tool and also depends on the Action component to correctly handle the input and output of the LLM in order to invoke external tools. Additionally, the Agent relies on the tool itself to provide the correct input and output to achieve accurate results. Therefore, issues such as inaccurate natural language descriptions of the tool, improper input/output formatting by the LLM, incorrect extraction of tool names and parameters by the Action component, non-compliant output parameters from the tool, and inherent problems within the tool itself can lead to unexpected behavior and prevent the Agent from successfully completing its task.

\section{Define the Defect of LLM-based Agent}
\label{sec:defect}

In this section, we conduct an empirical study of real-world posts on StackOverflow to define and categorize code defects in LLM-based Agents. StackOverflow is a developer Q\&A platform that allows developers to communicate and share solutions to software development issues they encounter~\cite{firouzi2024time}. In line with previous work~\cite{zhang2024demystifying,laoshiCordSort}, StackOverflow aggregates a large number of real-world programming problems, covering a wide range of languages, frameworks, and development scenarios~\cite{openja2020analysis}. Its advantages, such as tag classification and historical data, make it an ideal data source for collecting and categorizing defects in LLM-based Agents.

\subsection{Data Collection}

In this process, we collected posts related to defects in LLM-based Agents from StackOverflow. Specifically, we manually collected all posts up to August 2024 using two keywords: ``LLM Agent'' and ``AI Agent''. These searches yielded $378$ and $768$ posts, respectively. In addition, we collected posts related to four widely used Agent frameworks within the same time frame~\cite{sun2024cebench}, as many current Agents are built on top of these frameworks. The frameworks are Langchain~\cite{langchain}, AutoGen~\cite{autogen}, LlamaIndex~\cite{llamaindex}, and Flowise~\cite{Flowise}, from which we gathered $4,294$, $1,162$, $208$, and $44$ posts, respectively. In total, we collected $6,854$ posts.

\subsection{Data Pre-processing}

To filter out posts related to Agent code defects, we first screened the 6,854 collected posts, focusing only on those with explicitly accepted posts, as shown in Table~\ref{table:es}. 
To further refine the selection, we followed the approach outlined in prior research~\cite{ChenXLGLC22, zhang2024demystifying}. Three researchers, each with over three years of experience in AI and software development, participated in the data preprocessing. Two of them independently reviewed all $6,854$ posts, removing irrelevant and duplicate entries. In cases where their decisions conflicted, a third researcher was consulted, and the final decision was made through voting. The filtering criteria were as follows:
\begin{enumerate}
    \item Remove irrelevant posts; 
    \item Remove duplicate posts from previous ones; 
    \item Remove posts without an ``Accept Answer'' or no clear root cause to better define defects; 
    \item Remove posts that do not contain code; 
    \item Remove discussion posts, such as those comparing the capabilities of different LLMs. 
\end{enumerate}
As a result, $331$ posts were selected for further analysis.

\begin{table}[t]
\centering
\small
\caption{Posts collection and Card sorting process.}\label{table:es}
\scalebox{0.98}{
\begin{tabular}{l||c|c|c}
\hline \textbf{Keywords} &\textbf{Posts} & \textbf{Valid post} & \textbf{Cards} \\
\hline 

LLM Agent & 378 & 113 & 88\\
AI Agent & 768 & 35 & 15\\
LlamaIndex & 208 & 17 & 3\\
Flowise & 44 & 4 & 1\\
AutoGen & 1,162  & 12 & 3\\
Langchain & 4,294 & 487 & 221\\

\hline

\hline
\end{tabular}
}
\end{table}

\subsection{Data Analysis}
\begin{figure}[t]
	\centering{\includegraphics[scale=0.27]{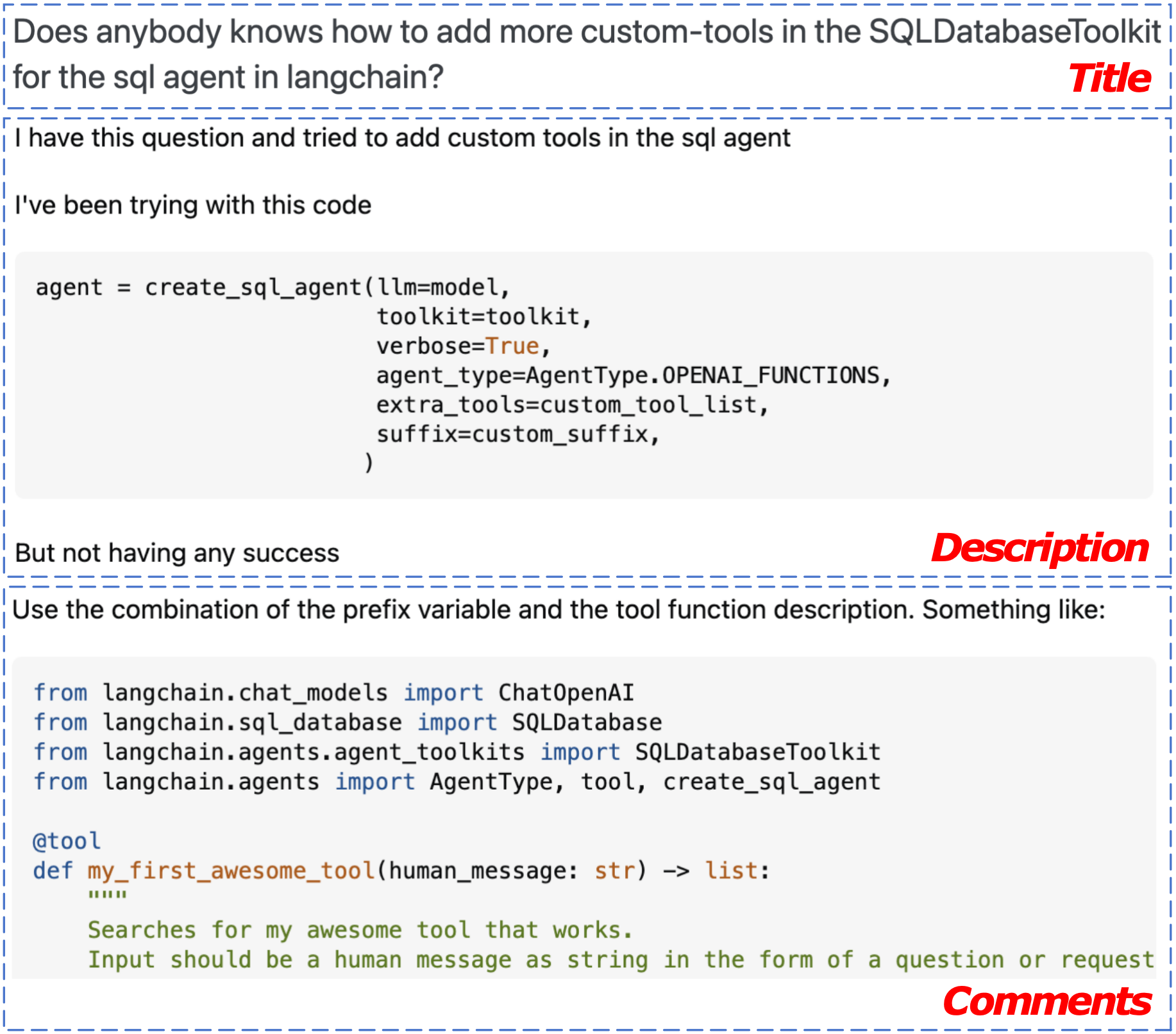}}
	\caption{An example of a card.}
	\label{Fig:cardexample}
\end{figure}

We manually analyzed the $331$ filtered posts to investigate the various types of code defects in LLM-based Agents. Since defects in LLM-based Agents have not yet been systematically studied, we did not use any predefined defect categories in this process. Instead, we adopted the open card sorting method~\cite{zampetti2021self,yang2024hyperion}, a commonly used information classification technique, to effectively define Agent code defect categories. Consistent with the preprocessing process, each report was represented by a card, which included a detailed description of the defect and its root cause. Fig.~\ref{Fig:cardexample} shows an example of a card from one of the posts. The card is divided into three parts: title, description, and comments. It describes a defect encountered by developers when using the Langchain project to develop Agents, where the LLM fails to correctly invoke external tools. The cause of this defect is the failure to accurately describe the functionality of the Tool and properly pass this information to the LLM. This leads to the Agent being unable to invoke the correct tool, resulting in unexpected behavior.

In the process of open card sorting, we ensure the representativeness of defects by considering the reproducibility of the code. Some issues may be specific to particular developers, such as an Agent failing to run correctly due to insufficient LLM API credits~\cite{chen2024empirical}. These issues will not be classified as representative defects. We follow the detailed steps from previous work~\cite{yang2023definition}. In the first round of classification, we randomly selected $40\%$ of the cards. We began by reading the title and description of each card to understand the issue it addresses. Then, we read the comments to understand the solutions and the root causes of the problems. Cards without a clear root cause were excluded, and potential defects were classified.

In the second round of classification, the two authors independently classified the remaining $60\%$ of the cards following the same steps described in the first round. Afterward, the authors compared their results, aligned the differences, and ultimately classified the defects into eight types, as shown in Table~\ref{table:defect}. It is important to note that some cards may correspond to multiple types of errors. For example, as shown in Fig.~\ref{fig::tv_example}.


\begin{table*}[t]
\centering
\small
\caption{Types and descriptions of defects.}\label{table:defect}
\scalebox{0.95}{
\begin{tabular}{l||l|l}
\hline \textbf{ID} & \textbf{Defect} & \textbf{Definition} \\
\hline 
ADAL & Adaptation Defect between Agent and LLM & The LLM used by the Agent fails to meet adaptability requirements.\\
IETI & Insufficient External Tool Information & Errors in tool registration information.\\
LOPE & LLM Output Parsing Error & LLM outputs do not meet requirements.\\
TRE & Tool Return Error & Tool return values are missing or contain implementation errors.\\
ALS & Action Listener Setting & Triggers set in the plan are unexpectedly activated.\\
MNFT & Missing Necessary Fault Tolerance & Essential fault tolerance is lacking in interactions with components like tools.\\
LARD & LLM API-related Defects & arise during LLM API calls.\\
EPDD & External Package Dependency Defect & Conflicts exist between external packages used by tools and the Agent.\\

\hline
\end{tabular}
}
\end{table*}

\subsection{Defects Definition}

Through our analysis of the card sorting, we identified eight types of defects in LLM-based Agents, covering all components of the Agent and its runtime processes as described in Section~\ref{sec2:Agent}, including model invocation, tool usage, plan execution, and short- and long-term memory retrieval. These defects can disrupt the normal operation of an Agent project, leading to crashes or unexpected behavior. Table 1 lists the names, IDs, and definitions of each defect type. In the following sections, we will provide a detailed explanation of each defect and illustrate them with examples.

\textbf{(1) Adaptation Defect between Agent and LLM (ADAL).} As the core component of an Agent, the LLM is responsible for executing complex operations such as invoking decision tools and retrieving external data based on user-designed reasoning chains and reflective prompts. For this, the LLM must possess strong natural language dialogue capabilities and general text generation abilities. 
However, some current LLMs are not well-suited for human dialogue tasks because they have not been fine-tuned for these applications, resulting in a lack of natural language conversation skills. Additionally, some LLMs are trained for specific tasks, such as CodeLLama~\cite{roziere2023code} (trained for code generation), or are outdated models that have not been upgraded, such as GPT-2~\cite{hanna2024does}. These LLMs often lack the necessary general text generation capabilities to correctly execute complex prompts, such as ReACT~\cite{cui2024receive}. Thus, this defect refers to cases where the LLM is unsuitable for use in an Agent due to the aforementioned reasons, which may prevent the Agent from functioning properly.

\textbf{Example:} Fig.~\ref{fig::aal_example} shows a real report from StackOverflow. In this example, the user employed StarCoder~\cite{li2023starcoder} as the Agent's LLM. Since StarCoder is a model trained specifically for code-related tasks, it was unable to handle the user’s prompt engineering, resulting in unexpected behavior. The Agent recursively called the LLM without terminating until the token length limit was reached, at which point the Agent returned an error.

\begin{figure}[h]
\centering
\begin{lstlisting}[language=Python]
class ChatLLM(BaseModel):
    ...
    def generate(self, prompt: str, stop: List[str] = None):
        response = client.chat.completions.create(
            model="file_path/StarCoder",
            messages=[{"role": "user", "content": prompt}],
            temperature=0.6)
        return response.choices[0].message.content
    \end{lstlisting}
\caption{An example of ADAL Defect.}
\label{fig::aal_example}
\end{figure}

\textbf{(2) Insufficient External Tool Information (IETI). }Agents extend the capabilities and knowledge of LLMs by preparing a series of external tools, such as calculators or weather information retrieval services. In an Agent system, these external tools must first be registered before being utilized. During tool registration, the tool’s name, description, and implementation code need to be provided. At runtime, the Agent selects the appropriate tool based on the registered information for different tasks. However, if there are errors in the registration, such as missing values or mismatches between the tool’s name, description, and implementation, the Agent may fail to correctly invoke the tool or invoke the wrong tool. This defect refers to problems in the registration information that result in the Agent’s failure to call tools, leading to abnormal behavior.

\textbf{Example:} Fig.~\ref{fig::ete_example} shows a post from StackOverflow regarding Langchain~\cite{langchain}. In this example, the Agent implemented a tool for linking to GitHub. However, both the name and description of the tool were left empty. As a result, the Agent was unable to utilize the tool and made an incorrect selection during tool invocation. This defect also occurs when there is a mismatch between the tool’s name, description, and actual implementation.

\begin{figure}[h]
\centering
\begin{lstlisting}[language=Python]
class GitHubAction(BaseTool):
    """Tool for interacting with the GitHub API."""
    api_wrapper: GitHubAPIWrapper = Field(default_factory=GitHubAPIWrapper) 
    mode: str
    name: str = ""
    description: str = ""
    args_schema: Optional[Type[BaseModel]] = None
    def _run(
    ...
    \end{lstlisting}
\caption{An example of IETI Defect.}
\label{fig::ete_example}
\end{figure}

\textbf{(3) LLM Output Parsing Error (LOPE).} When generating content, an LLM may produce outputs that do not meet the intended requirements due to various reasons, such as hallucinations~\cite{ye2023cognitive}. 
During runtime, the Agent must continually parse the LLM’s output to proceed with subsequent operations. Therefore, incorporating necessary fault tolerance during output parsing is essential to ensure the proper functioning of the Agent. This defect refers to the absence of fault-tolerant mechanisms in the Agent’s implementation for handling model outputs, which can lead to unexpected behavior in the Agent.

\textbf{Example:} Fig.~\ref{fig::lpe_example} shows an example from StackOverflow regarding Langchain. In this case, the user set \textit{handle\_parsing\_error} to False while using the Agent, which resulted in the model’s output not meeting expectations and being unparseable. This caused the Agent project to terminate unexpectedly.

\begin{figure}[h]
\centering
\begin{lstlisting}[language=Python]
agent = initialize_agent(
    tools = tools,
    llm = llmm,
    agent = "zero-shot-react-description",
    verbose = True, 
    handle_parsing_error = False)
    \end{lstlisting}
\caption{An example of LOPE Defect.}
\label{fig::lpe_example}
\end{figure}

\textbf{(4) Tool Return Error (TRE).} After the Agent successfully commands the LLM to invoke an external tool, it must wait for the tool to execute and return results to the LLM. However, tools may involve chained function calls during their implementation, and if any function lacks a return value or if there is a defect in the tool’s implementation, the tool may fail to return a result. This can cause the Agent to halt at that point, leading to service interruptions. Therefore, this defect refers to cases where any stage of the tool’s implementation lacks a return value or has implementation defects, causing the Agent’s service to be interrupted.

\textbf{Example:} Fig.~\ref{fig::tre_example} shows an example from StackOverflow related to Langroid~\cite{liu2024demystifying}. In this case, the Agent implemented a tool called \textit{SegmentExtractTool}. However, one of the essential functions, instructions, lacked a return value, causing the Agent’s service to be interrupted. Similar defects can occur when tools are improperly implemented or when the tool implementation is incomplete.

\begin{figure}[h]
\centering
\begin{lstlisting}[language=Python]
class SegmentExtractTool(ToolMessage):
    ... 
    @classmethod
    def instructions(cls) -> str:
        return 
        """
        Use this tool/function to indicate certain segments from a body of text containing numbered segments.
        """
    \end{lstlisting}
\caption{An example of TRE Defect.}
\label{fig::tre_example}
\end{figure}

\textbf{(5) Action Listener Setting (ALS). }Agents control the model’s output by specifying trigger words in the prompt, using these trigger words to decide the next action, such as invoking tools or terminating the model’s output. However, if external inputs, such as the return value from a tool, contain the trigger word, or if certain parameters within the Agent (e.g., tool names) are passed into the prompt and happen to include the trigger word, the Agent may prematurely detect the trigger word. This can result in the Agent ending the service prematurely. This defect is particularly common in cases where trigger words are set in a generic manner, such as the default trigger word ``None'' in Langchain.

\textbf{Example:} Consider the \textit{run} function of the Agent in Fig.~\ref{fig::tv_example}. Typically, the Agent checks the tool’s name to determine when to invoke or stop using a tool and inserts trigger words to control the model’s decisions. However, several instances of ALS defects are present in this case. First, if a tool’s name is set to ``Final,'' the Agent incorrectly interprets this as an indication that the service has ended, skipping the tool invocation and directly producing an erroneous output, resulting in unexpected behavior (line 4). Second, if the tool’s return value matches the trigger word (e.g., ``observation''), the presence of two identical trigger words can cause the LLM to enter a loop, leading to incorrect decision-making and unexpected outcomes (line 8).

\begin{figure}[h]
\centering
\begin{lstlisting}[language=Python]
def run(self, question: str):
    while num_loops < self.max_loops:
        ...
        if tool == 'Final':
            return tool_input
        assert isinstance(tool_input, str)
        tool_result = self.tool_by_names[tool].use(tool_input)
        generated += f"\n{OBSERVATION_TOKEN} {tool_result}\n{THOUGHT_TOKEN}"
        previous_responses.append(generated)
    \end{lstlisting}
\caption{An example of ALS and MNFT Defect.}
\label{fig::tv_example}
\end{figure}

\textbf{(6) Missing Necessary Fault Tolerance (MNFT). }As described in Section 2.1, Agents frequently interact with external tools or memory during their operation. Moreover, the Agent’s workflow is controlled by both natural language and code, leading to low coupling between components. Therefore, it is essential for the Agent to implement necessary fault tolerance when invoking external tools or memory to ensure the service runs smoothly. This defect occurs when the Agent fails to implement proper fault tolerance for the input and output of external tools or data invocations, resulting in service interruptions.

\textbf{Example:} In the example from Fig.~\ref{fig::tv_example}, the Agent also exhibited an MNFT defect. Before invoking the tool, the Agent did not check whether the tool existed or validate the type of the value being passed to the tool (line 7). Additionally, no fault tolerance checks were performed on the tool’s output. Typically, the LLM only accepts strings as input, and if the tool returns a value of a different type, passing it directly to the LLM would cause an error, leading to a service interruption in the Agent.

\textbf{(7) LLM API-related Defect (LARD).} During runtime, Agents may make various API calls, such as invoking an LLM or external links through tools. This defect refers to defects that arise when the Agent encounters problems during API calls, which can lead to service interruptions. Such defects are often concentrated in situations where the Agent is handling LLM-related API calls.

\textbf{Example:} In the example from Fig.~\ref{fig::ARD_example}, the ChatLLM class needed to invoke an API provided by OpenAI to call the LLM. However, since the \textit{api\_key} was empty, ChatLLM failed to correctly invoke the LLM (line 2). Additionally, the API call did not specify a trigger word, which prevented the Agent from properly terminating the reasoning process and invoking external tools. As a result, the Agent failed to function correctly (line 5).

\begin{figure}[h]
\centering
\begin{lstlisting}[language=Python]
class ChatLLM(BaseModel):
    api_key:str = ""
    def generate(self, prompt: str, stop: List[str] = None):
        client:OpenAI = OpenAI(api_key=self.api_key, base_url=url)
        response = client.chat.completions.create(
            model='gpt-4o',
            messages=[{"role": "user", "content": prompt}] )
        return response.choices[0].message.content
    \end{lstlisting}
\caption{An example of LARD Defect.}
\label{fig::ARD_example}
\end{figure}

\textbf{(8) External Package Dependency Defect (EPDD).} The implementation of tools in an Agent is decoupled from the Agent framework, allowing users to implement and register tools according to their specific needs. 
However, if there is an overlap between the external packages that the Agent relies on and those required by the framework, version conflicts may arise. These conflicts can lead to interruptions in the Agent’s services. Such conflicts can lead to service interruptions for the Agent.

\textbf{Example: }In the Agent project shown in Fig.~\ref{fig::epdc_example}, both the Agent framework and an external tool, \textit{Python\_repl}, depend on the pydantic package. If there is a version mismatch between the pydantic package versions required by the framework and the tool, this can lead to service interruptions or prevent the Agent from functioning correctly when invoking the tool.

\begin{figure}[h]
\centering
\begin{lstlisting}[language=Python]
""" File: Path/Agent.py"""
from pydantic import BaseModel
class Agent(BaseModel):

""" File: Path/Tools/Python_repl.py"""
from pydantic import BaseModel
class PythonREPL(BaseModel):
    \end{lstlisting}
\caption{An example of EPDD Defect.}
\label{fig::epdc_example}
\end{figure}

\textbf{Defect vs. Bug vs. Vulnerability.} We use the term ``defect'' to collectively refer to issues present in the Agent project. Compared to other terms such as ``vulnerability'' or ``bug'', ``defect'' is broader and more representative of these problems. Specifically, a vulnerability refers to a defect that can be directly exploited but does not encompass potential threats. For instance, MNFT does not necessarily lead to a service interruption in the Agent; only in certain situations, as illustrated in Fig.~\ref{fig::ARD_example}, does it cause this issue. Additionally, a ``bug'' refers to a defect caused by coding errors. However, defects like AAL are often the result of design flaws, where the Agent and LLM are not well-adapted, causing the Agent to malfunction—this is not a coding error.

\section{Detect the Defect of LLM-based Agent}
\label{sec:detect}

The results presented in Section~\ref{sec:defect} identified eight types of defects in Agent projects. To provide real-world evidence of these defects and assist developers in detecting them in practice, we developed Agentable, an automated testing tool designed to detect the defined eight defects in Agent projects.

\subsection{Challenges and Design Decisions}

When detecting defects in Agent projects, the complexity of the real-world environment in which the Agent operates and the uncertainty of the Agent development model introduce new challenges~\cite{wang2024large,deng2024ai,zhang2024privacyasst}. In the following sections, we will discuss these challenges and describe the design decisions we made to address them.

\textbf{Complex Defect Triggering Patterns.} The workflow of an Agent is controlled by both code and natural language, which involves complex execution contexts. For example, in the ALS defect, the tool’s return value, the user’s prompts, and predefined trigger words can all contribute to the occurrence of this defect. This type of defect relies not only on control flow but also on the semantics of the code and data dependencies. As a result, traditional data flow-based analysis is insufficient to detect such errors. Detecting this type of defect requires multi-dimensional feature extraction from the code. Furthermore, real-world Agent implementations do not follow a unified framework or pattern, which further complicates the vulnerability model of Agents.

\textbf{Design Decision.} To address this challenge, Agentable constructs a Code Property Graph (CPG) to extract vulnerability features related to defects. A CPG provides a joint representation of the syntax, control flow, and data flow of the target code, capturing more comprehensive code features. Based on the CPG, Agentable designs distinct patterns for each type of vulnerability to locate defects effectively. Additionally, since CPGs are often complex and may exhibit significant syntactic variation under the same semantics, Agentable introduces appropriate abstraction in its node queries to handle these variations.


\textbf{Complex Logical Defects. }Agent defects often include logical flaws, which can be triggered through natural language inputs and may not result in runtime errors but instead manifest as behavior inconsistent with expectations. For example, in the ETE defect, the mismatch between the tool’s description, name, and implementation triggers the defect. However, the triggering conditions may arise not only from logical defects within the tool itself but also from inaccurate tool descriptions. Moreover, logical defects are often characterized by their strong generalization properties, making them difficult to detect solely by identifying code elements. Thus, traditional detection approaches that rely only on code elements struggle to identify such Agent defects.

\textbf{Design Decision.} To address this challenge, Agentable incorporates LLMs into the generalized search and logic reasoning components to enhance the accuracy of defect detection. For instance, LLMs are used to generate code summaries for tools and evaluate whether the tool’s implementation aligns with its description. To further improve accuracy, we designed prompt patterns based on chain-of-thought techniques for each stage of the generalized search and logical reasoning. By leveraging the powerful code comprehension capabilities of LLMs, Agentable can better identify potential logical defects in Agent projects.


\textbf{Defects Dependent on Uncertain State.} Some defects are triggered by the generated output of the LLM and the return values from external tools. For example, in the case of an LPE defect, whether the defect is triggered depends entirely on whether the format of the LLM’s output is correct. These types of defects are dependent on the state during the Agent’s execution process. In such cases, dynamic testing methods, such as fuzz testing, cannot guarantee that all possible outputs from the LLM will be tested. Therefore, dynamic testing alone is insufficient to detect these defects. An effective way to address these defects is through detailed semantic analysis. However, when using Joern to construct the CPG, some semantic information is lost~\cite{chen2024utilizing,bui2023detecting}, making it even more challenging to detect such defects.

\textbf{Design Decision.} To address this challenge, Agentable enhances the CPG query results by locating the corresponding source code and constructing a semantic enrichment module based on the Abstract Syntax Tree (AST). This module restores the semantic information missing from the CPG, such as data dependencies related to LLM inputs. Leveraging these enriched semantics, Agentable employs a series of defect-specific detection strategies to effectively identify potential defect-triggering patterns.

\subsection{Overview}

\begin{figure*}[h]
	\centering{\includegraphics[scale=0.43]{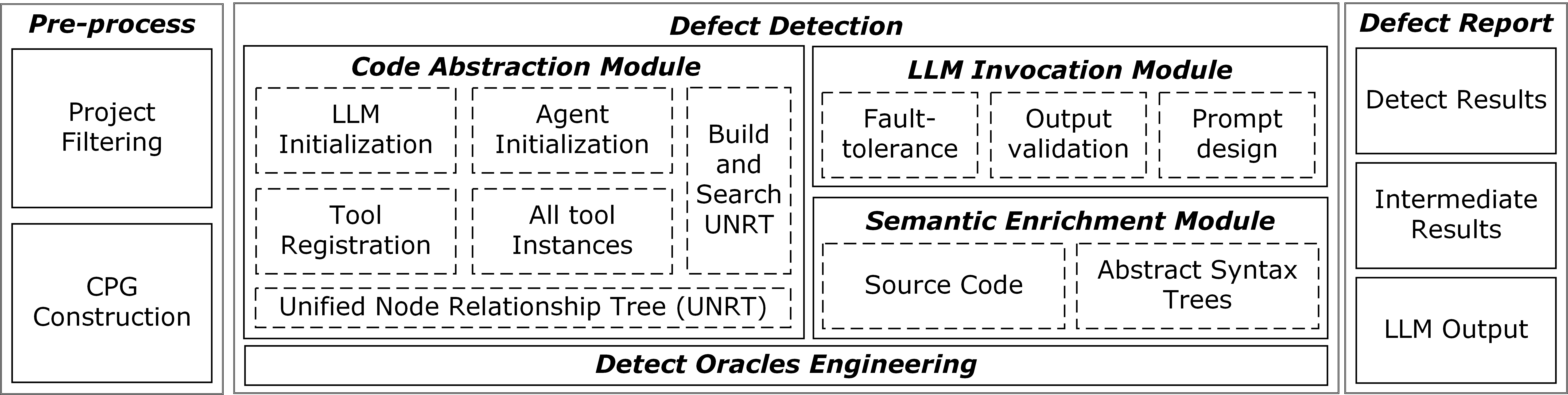}}
	\caption{The overview of Agentable.}
	\label{Fig:overview}
\end{figure*}

Fig.~\ref{Fig:overview} provides an overview of Agentable’s workflow. Given an Agent project for analysis, Agentable first preprocesses the project, filtering out irrelevant information to more accurately construct the project’s CPG. Next, Agentable enters the defect detection phase, where it invokes the three sub-modules in the defect detection module according to the patterns of different defects: code sections requiring localization are handled by the CPG-based code abstraction module, tasks requiring generalized search or logical reasoning are managed by the LLM invocation module, and parts requiring semantic enhancement are processed by the semantic enrichment module. Finally, Agentable analyzes the output of the defect detection module in the defect summary phase, identifies the defects present in the Agent project, and generates a defect report.

\subsection{Pre-process}

There are two phases in the preprocessing phase, i.e., Project Filtering and CPG Construction. Agentable first filters the files in the Agent project to ensure that the project’s CPG can be constructed quickly and accurately.

\textbf{Project Filtering.} Agentable retains only code-related files and removes irrelevant files, such as documentation. The purpose of this is twofold: first, to accelerate the construction of the project’s CPG, and second, to prevent irrelevant files from affecting the accuracy of the CPG construction, which could impact the subsequent detection results.

\textbf{CPG Construction.} Agentable leverages Joern, a widely used open-source code analysis tool, to construct the CPG for the project. Agentable takes the root directory of the filtered target Agent project, constructs its CPG, and saves the graph representation of the CPG into node and edge files.

\subsection{Defect Detection}

In the defect detection phase, we first provide an overview of the three modules used for defect detection, explaining their working principles and scope. Next, we present the detection patterns for each specific defect in detail.

\subsubsection{Submodular Design} In this part, we provide a detailed explanation of the three modules involved in the defect detection process of Agentable: the Code Abstraction Module, the LLM Invocation Module, and the Semantic Enrichment Module.
\leavevmode\newline\textbf{Code Abstraction Module.} In the defect detection process, parts of the code that require localization are handled by the code abstraction module. Among the eight types of defects, we primarily need to locate the following four types of code implementations: \ding{182} the initialization of the LLM, \ding{183} the initialization of the Agent, \ding{184} the initialization of the Tool, and \ding{185} all instances of the Tool. Since real-world Agents may implement similar semantics in significantly different ways, Agentable applies abstraction during the CPG query process. Specifically, it treats functions and classes as equivalent nodes (referred to as unified nodes) but preserves distinctions in the relationships between classes (inheritance), functions (function calls), and the encapsulation relationship between classes and their contained functions. Agentable first queries all unified nodes in the CPG and retains the three types of relationships between them. Next, it constructs a unified node relationship tree (UNRT) with the process described in Algorithm~\ref{Alg:construct}.

\begin{enumerate}
    \item First, Agentable extracts all classes and constructs a class inheritance tree based on their inheritance relationships, forming the trunk. 
    \item Next, Agentable extracts all functions encapsulated within each class and attaches them to the corresponding class on the trunk, forming branches. 
    \item Furthermore, Agentable supplements the functions called by the functions encapsulated within the class (e.g., if function A calls function B, function B is attached after function A), forming leaves. It is important to note that if a function is called by functions from multiple classes, that node will be added to multiple branches.
    \item Finally, Agentable similarly adds the functions that call the encapsulated functions (e.g., if function A is called by function B, function B is attached after function A), forming additional leaves. Likewise, leaves that are called multiple times are duplicated accordingly.
\end{enumerate}

\begin{algorithm}[t]
\footnotesize
\caption{Unified Node Relationship Tree (UNRT) Construction Process}
\label{Alg:construct}
\begin{algorithmic}[1]
    \STATE \textbf{Input:} CPG from Joern
    \STATE \textbf{Output:} UNRT
    \STATE Initialize an empty class inheritance tree $\mathcal{T}$
    \FOR{each class $C_i$ in the source code}
        \STATE Add $C_i$ to $\mathcal{T}$ with its inheritance relationships as edges
    \ENDFOR

    \FOR{each class $C_i$ in $\mathcal{T}$}
        \FOR{each function $f_{j}$ in $C_i$}
            \STATE Attach $f_{j}$ to $C_i$ as a branch node
        \ENDFOR
    \ENDFOR

    \FOR{each function $f_{j}$ in each class $C_i$}
        \FOR{each function $f_k$ called by $f_{j}$}
            \STATE Attach $f_k$ after $f_j$ as a leaf node
            \IF{$f_k$ is called by functions in multiple classes}
                \STATE Duplicate $f_k$ for each call in the respective branches
            \ENDIF
        \ENDFOR
    \ENDFOR

    \FOR{each function $f_{j}$ in each class $C_i$}
        \FOR{each function $f_m$ that calls $f_{j}$}
            \STATE Attach $f_m$ after $f_j$ as an additional leaf node
            \IF{$f_m$ calls $f_j$ multiple times}
                \STATE Duplicate $f_m$ accordingly
            \ENDIF
        \ENDFOR
    \ENDFOR
\end{algorithmic}
\end{algorithm}

Subsequently, we locate the required code implementations by searching through UNRT. The overall search strategy is outlined in Algorithm~\ref{Alg:search}. First, we search layer by layer along the trunk until a node that meets the criteria is found. If no suitable node is found after searching the entire trunk, we proceed to search the branches, again layer by layer, until a matching node is located. If no result is found, we continue by searching the leaf nodes layer by layer until the search is complete. By constructing a unified node relationship tree, Agentable not only accelerates the search process but also ensures compatibility with different implementations of Agents that share the same semantics.

\textbf{LLM Invocation Module.} In the detection process, all tasks requiring generalization, logic, or semantic reasoning are handled by the LLM invocation module. First, to enhance the accuracy of the LLM’s judgments, we designed prompts based on the chain-of-thought technique. Second, to facilitate extraction from the LLM’s output, we strictly controlled the format of the LLM’s responses and specified a fixed output format. Finally, to prevent potential negative impacts from the LLM’s output, we incorporated fault-tolerance handling and format validation modules both before and after invocation, increasing the accuracy of the LLM’s results.

\textbf{Semantic Enrichment Module.} In the detection process, all tasks requiring detailed semantic information are handled by the semantic enrichment module. We integrate semantic information from two sources: the source code and the corresponding AST. The source code helps recover runtime state judgments that are missing from the CPG, while the AST restores missing information, such as the types of class attribute initialization values. By enriching the semantics from these two aspects, we can uncover more potential insights from the source code, such as risks related to runtime conditions.

\begin{algorithm}[h]
\footnotesize
\caption{Layered Search on UNRT}
\label{Alg:search}
\begin{algorithmic}[1]
    \STATE \textbf{Input:} UNRT $\mathcal{T}$
    \STATE \textbf{Output:} Node matching results
    \STATE Initialize search result as \texttt{None}
    \FOR{each part $P$ in \{trunk, branches, leaves\} of $\mathcal{T}$}
        \FOR{each layer $L$ in $P$}
            \FOR{each node $n$ in layer $L$}
                \IF{$n$ meets the criteria}
                    \STATE Set search result to $n$ and \textbf{exit} search
                \ENDIF
            \ENDFOR
        \ENDFOR
        \IF{search result is not \texttt{None}} 
        \STATE \textbf{break}
        \ENDIF
    \ENDFOR
    \STATE \textbf{Return} search result
\end{algorithmic}
\end{algorithm}

\subsubsection{Detect Oracles} In this part, Agentable executes defect detection patterns to uncover potential vulnerabilities in the Agent project. Below, we provide a detailed description of these detection Oracles and explain how each Oracle invokes the three modules in the detection process.

\textbf{(1) Adaptation Defect between Agent and LLM (ADAL).} Agentable first checks the parts of the Agent project related to the initialization of the LLM to identify potential AAL defects. Specifically, Agentable begins by invoking the code abstraction module and using Algorithm~\ref{Alg:search} to traverse the Unified Node Relationship Tree (UNRT) to search for LLM initialization or Agent initialization nodes, as the LLM’s initialization code may be embedded within the Agent’s initialization. For each node, Agentable then calls the semantic enrichment module to retrieve the source code and its corresponding AST, grouping every $n$ node and its source code together. Subsequently, Agentable invokes the LLM invocation module to evaluate, group by group, whether the nodes correspond to LLM or Agent initialization nodes. If such a node is found, the search is terminated, and relevant LLM information is extracted. Finally, based on the LLM’s name and the Hugging Face interface, Agentable determines whether the LLM used by the Agent is a general-purpose model or a dialogue model. If the LLM is a task-specific model or not a dialogue model, Agentable reports an AAL defect. The detection results and detailed information are then passed to the defect summary module.

\textbf{(2) Insufficient External Tool Information (IETI). }Agentable checks all instances of tools in the Agent project to detect potential ETE defects. Specifically, Agentable first invokes the code abstraction module and uses Algorithm~\ref{Alg:search} to traverse all nodes to locate Tool initialization nodes. The method for determining whether a node is a Tool initialization node is the same as for AAL detection: nodes are grouped in sets of $n$ and searched layer by layer from the trunk to the leaves. Agentable then calls the semantic enrichment module to extract the source code and invokes the LLM invocation module for further evaluation. Once a Tool initialization node is identified, we query its child nodes (i.e., all nodes following that point on the trunk, branches, and leaves) to find all Tool instances. Agentable then uses the semantic enrichment module to extract the registration information and implementation code for each Tool instance.

Next, Agentable invokes the LLM invocation module, inputting both the Tool’s registration information and implementation code, to evaluate two key questions: \ding{182} Whether there is a discrepancy or missing value between the Tool’s registration information (name, description) and its implementation code. \ding{183} Whether there are discrepancies or missing values within the Tool’s registration information itself. If either of these defects is detected, Agentable reports an ETE defect. The detection results and detailed information are then passed to the defect summary module.

\textbf{(3) LLM Output Parsing Error (LOPE).} Agentable checks all nodes in the Agent project that invoke the LLM and verifies whether the inputs and outputs of these invocations include necessary fault tolerance within their context. Specifically, Agentable first searches the Unified Node Relationship Tree (UNRT) to locate Agent initialization nodes. The method for determining whether a node is an Agent initialization node follows the same process as for AAL detection: nodes are grouped in sets of 10 and searched layer by layer from the trunk to the leaves. Agentable then calls the semantic enrichment module to extract the source code and invokes the LLM invocation module for evaluation.

Next, based on the LLM initialization nodes and code identified during AAL detection, Agentable uses the LLM invocation module to identify the specific function names executed during LLM initialization, such as generate or create, which are commonly invoked by the Agent when calling the LLM. Based on the results from the LLM invocation module, Agentable calls the semantic enrichment module to extract the code context surrounding the Agent’s LLM invocations. The context includes the scope of inputs and outputs for the LLM invocation.

Since there is no universal standard for fault tolerance, common fault tolerance mechanisms include constructs like “try-except” and “assert.” Thus, the extracted code context is also input into the LLM invocation module to determine whether fault tolerance mechanisms are present for both the inputs and outputs of the LLM invocation. If either the input or output lacks fault tolerance, Agentable reports an ETE defect. Finally, Agentable outputs the detection results and detailed information to the defect summary module.

\textbf{(4) Tool Return Error (TRE).} Agentable checks all tool instances in the Agent project and verifies the return values and implementations of all functions within the tools. Specifically, Agentable reuses the results from the ETE detection phase, which includes the implementation code of all tool instances. Next, Agentable invokes the semantic enrichment module to retrieve the ASTs of different tool instances, allowing it to check all functions within each tool and their return values. For each tool instance, Agentable conducts the following two checks: \ding{182} it verifies whether all functions within the tool have return values, \ding{183} it uses the LLM invocation module to check the tool’s code for potential defects. If any problem is detected in either of these two checks, Agentable reports a TRE defect. Finally, Agentable outputs the detection results and detailed information.

\textbf{(5) Action Listener Setting (ALS). }Agentable checks whether there are conflicts between the trigger words in the Agent project and the names or return values of tools. Specifically, Agentable first reuses the intermediate results from the LPE detection phase to obtain the Agent initialization nodes. Next, Agentable searches for the list of trigger words from these nodes and their child nodes by examining the stop lists flowing into the LLM and extracting the trigger words. Since trigger words may also include variables, Agentable additionally invokes the semantic enrichment module to recursively backfill intermediate variables into the trigger words, generating the actual trigger word list.

Furthermore, Agentable retrieves the intermediate results from the ETE detection phase to obtain the code for all tool instances. For each tool, Agentable invokes the semantic enrichment module to extract the tool’s return values and name and compares them against the trigger word list. If a tool’s return value or name is found within the trigger word list, Agentable reports a TRE defect. Agentable outputs the detection results and relevant intermediate information.

\textbf{(6) Missing Necessary Fault Tolerance (MNFT). }Agentable checks whether the inputs and outputs of all nodes that invoke tools in the Agent project include necessary fault tolerance within their context. Specifically, Agentable first reuses the intermediate results from the LPE and ETE detection phases to obtain the initialization nodes for both the Agent and the tools. Next, Agentable invokes the LLM invocation module to identify the actual function names executed at the Tool initialization nodes, such as run or invoke. Then, Agentable uses the semantic enrichment module to extract the inputs, outputs, and their respective scopes when the Agent invokes the Tool. Finally, Agentable calls the LLM invocation module to determine whether the inputs and outputs in these scopes have the necessary fault tolerance. If they do not, Agentable reports an MNFT defect.

\textbf{(7) LLM API-related Defect (LARD). }Agentable checks the method of invoking the LLM in the Agent project and ensures the correctness of the invocation pattern. Specifically, Agentable first reuses the intermediate results from AAL detection to locate the LLM initialization class, and also reuses the intermediate results from LPE to identify the specific function name (function C) executed within the LLM initialization code. Next, Agentable invokes the semantic enrichment module to analyze the data flow into function C across all functions in the Agent class, extracting the relevant code involved in the data flow.

Then, Agentable uses the LLM invocation module to analyze both the Agent’s initialization code and the extracted data flow-related code, checking for errors in the LLM invocation. This includes verifying whether necessary values like api\_key are properly initialized or whether the invocation is incorrect (e.g., missing essential parameters like stop). If Agentable detects uninitialized necessary values or incorrect invocation patterns, it reports an ARD defect.

\textbf{(8) External Package Dependency Defect (EPDD). }Agentable checks whether there is any overlap between the external packages used by the Tool and those used by the Agent. Specifically, Agentable first reuses the results from IETI detection to obtain all Tool instances. For each Tool instance, Agentable first invokes the semantic enrichment module to analyze the external packages used by the Tool instance. Next, Agentable calls the semantic enrichment module again to analyze the external packages used by the Agent after excluding the specific Tool instance. Finally, Agentable compares the packages used by both the Tool and the Agent to check for any overlap. If overlapping packages are detected, Agentable reports an EPDD warning.

\subsection{Defect Report}
At this stage, Agentable combines the results from the defect detection process along with some intermediate steps, such as the formatted evaluation results from the LLM invocation module, to generate a comprehensive detection report. The detection report is then output to a specified folder. The generated report not only indicates whether each component of an Agent contains defects, but also provides detailed descriptions of the identified defects. These descriptions help developers identify the underlying causes of the defects and offer guidance on how to resolve them effectively.

\section{Evaluation}
\label{sec:evaluation}

The evaluation has two main objectives. First, We constructed two datasets, AgentSet and AgentTest, to evaluate the effectiveness and reliability of Agentable’s defect detection. AgentSet contains $84$ real-world Agents, which are used to assess the effectiveness of Agentable’s defect detection. AgentTest, on the other hand, contains $78$ manually constructed Agents, each with a single defect, and is used to evaluate the reliability of Agentable’s defect detection. Second, by analyzing the results of the experiment, we uncover the eight types of defects present in Agents deployed in real-world environments and gain a deeper understanding of their prevalence and distribution.

\subsection{Evaluation Setup}

\textbf{Dataset Construction of AgentSet. }We constructed a real-world dataset comprising $84$ Agent projects called AgentSet to evaluate the usability of Agentable. We searched for open-source projects on GitHub using the keywords ``LLM Agent'' and ``AI Agent''. As of August, 2024, these keywords returned $2.6k$ and $11.4k$ results, respectively. We manually filtered the results according to the following criteria: 
\begin{itemize}[leftmargin=1em]
    \item Exclude irrelevant results such as Agent Paper Lists or Agent Benchmarks; 
    \item Exclude empty repositories, repositories with no code, or duplicate repositories; 
    \item Exclude Agent projects not primarily implemented in Python; 
    \item Exclude projects that do not fit the definition of an LLM-based Agent used in this paper, such as those unable to invoke tools or Multi-Agent systems;
    \item To ensure the quality of the Agents, only retain projects with more than $200$ stars.
\end{itemize}
After filtering, we obtained $61$ and $23$ Agent projects for the two keywords, respectively. In total, we obtained $84$ Agent projects to AgentSet. These Agents exhibit significant differences in development models and code structures, making them effective for evaluating Agentable’s defect detection capabilities.

\textbf{Dataset Construction of AgentTest. }We constructed AgentTest to evaluate the recall of Agentable’s defect detection, which requires a dataset containing both positive and negative cases. To reflect real-world scenarios, we built the construction upon the $LLM\_Agents$ project~\cite{basellm}. This project was selected because it aligns with the definition of an Agent and is more adaptable to modification. First, we conducted a thorough manual review of the project and confirmed that it contained one LPE defect, one MNFT defect, and one LARD defect. Next, we manually fixed these defects and ran Agentable for detection, confirming that no defects were present. Furthermore, based on the detection results of RQ1, we constructed specific Agent scenarios containing individual defects, ensuring that the total number still met the $95\%$ confidence level and a confidence interval of $10$. Each Agent contains only one defect. Finally, we allocated the construction distribution of defect Agents according to the distribution of results in RQ1 and rounded up decimal values to ensure broader coverage of defect Agents. The detailed results are shown in Table~\ref{table:rq2}.

\textbf{Agentable Implementation and Setup. }Agentable is implemented in Python, with some embedded Scala statements to query the CPG generated by Joern~\cite{lee2023adcpg}. We implement Agentable with around $9,000$ lines of code in Python. For the LLM selection, Agentable uses GPT-4o-mini~\cite{isogai2024toward} due to its strong code processing capabilities and cost-effectiveness. To prevent context overflow in the LLM, we set n = $10$. The AST parsing in the semantic enrichment module is based on Python’s built-in AST package. All experiments are conducted on a workstation with $128$ CPU cores, $8$ $\times$ NVIDIA $A800$ ($80$G) GPUs, and Ubuntu 20.04.4 OS.

\subsection{Research Questions}
Specifically, we focus on the following three research questions.

\begin{enumerate}[label=\textbf{RQ\arabic*.},leftmargin=0.95cm]
    \item  How does Agentable perform on the AgentSet dataset? Can it detect defects with high precision?
    \item  What is the recall rate of Agentable in identifying agent-related defects?
    \item  What is the prevalence and distribution of defects in real-world, field-deployed agents?
\end{enumerate}

\subsection{RQ1: Detecting Defects in the Large-Scale Dataset}

To answer RQ1, we ran Agentable on AgentSet and analyzed the results. Agentable took $68$ hours to analyze these Agents, with an average analysis time of $0.81$ hours per project. In total, Agentable reported $889$ defects. Table~\ref{table:rq1} shows a detailed breakdown of Agentable’s analysis results for each defect type.

\begin{table}[h]
\centering
\small
\caption{Defects detected by Agentable.}\label{table:rq1}
\scalebox{0.98}{
\begin{tabular}{l||c|c|c|c|c}
\hline \textbf{Defect} &\textbf{Detected} & \textbf{Sampled} & \textbf{TP} & \textbf{FP} & \textbf{Precision} \\
\hline 

ADAL & 76 & 43 & 40 &3 & 93.02\%\\
IETI & 110 & 52 & 47 & 5& 90.38\%\\
LOPE & 56 & 36 & 27 & 9&75\%\\
TRE & 140 & 57 & 46 & 11& 80.7\%\\
ALS & 2 & 2 & 1  & 1 & 50.0\%\\
MNFT & 21 & 17 & 16 & 1& 94.12\%\\
LARD & 139 & 57 &49  &8 & 85.96\%\\
EPDD & 345 & 75 & 75 &0 &100.0\%\\
\hline
Total & 889 & 339 & 301 &38 &88.79\%\\

\hline

\hline
\end{tabular}
}
\end{table}

\textbf{Accuracy.} To evaluate the precision of Agentable in detecting each type of defect, we manually analyzed the defects reported by Agentable in the experiment. Consistent with prior research~\cite{zhang2024demystifying,yang2024hyperion,lee2023adcpg}, 
we randomly sampled a number of defects for each defect type to make the manual analysis feasible. The sample size for each defect type was carefully chosen to achieve a confidence level of $95\%$ and a confidence interval of $10$. The second and third columns of Table~\ref{table:rq1} show the number of Agent defect reports detected and sampled for each defect type.

Subsequently, two authors independently labeled these Agent defect reports as either true positives (TPs) or false positives (FPs), with a third author resolving any disagreements when necessary. The fourth to sixth columns of Table~\ref{table:rq1} list the number of true positives, false positives, and precision rates for each defect type, respectively. We then calculated the overall precision of Agentable as the weighted average of these precision rates, with the weight corresponding to the number of defects for each type. As a result, Agentable achieved an overall precision of $88.79\%$.

\textbf{False Positives.} Upon reviewing the false positives reported by Agentable, we identified two primary factors contributing to these errors. The first is the presence of overly abstracted classes in real-world Agent projects. For example, in detecting MNFT defects, we first need to identify the Agent’s initialization class and the Tool’s initialization class. The defect is then determined by examining the context in which the Tool is invoked in the Agent’s initialization class. However, we found that in some Agent projects, such as Langchain, there are numerous abstractions and nestings based on BaseModel, making it difficult to effectively identify the Agent’s initialization class, thus leading to detection failures.

The second major cause is the multiple control schemes available in some Agent projects. Some Agent projects, particularly Agent frameworks, provide various control mechanisms during construction. This can lead to Agentable mistakenly reporting defects for control schemes that were not selected.

\textbf{Overhead.} The total cost of Agentable for LLM invocations was $\$24.2$, with an average cost of less than $\$1$ per Agent. On average, each defect detection incurred a cost of approximately $\$0.027$.

\subsection{RQ2: Evaluating Agentable on the Annotated Dataset}

To answer RQ2, we applied Agentable to the AgentTest dataset to evaluate the reliability of Agentable’s defect detection. The detection results are shown in Table~\ref{table:rq2}.

\begin{table}[h]
\centering
\small
\caption{Defect-Agents detected by Agentable.}\label{table:rq2}
\scalebox{0.98}{
\begin{tabular}{l||c|c|c|c}
\hline \textbf{Defect} &\textbf{defect Agents} & \textbf{TP} & \textbf{FP} & \textbf{Precision} \\
\hline 

ADAL & 10 & 10 &0 & 100\%\\
IETI  &12 & 10 & 2& 83.33\%\\
LOPE &8 & 7 & 1& 87.5\%\\
TRE& 13 & 10 & 3& 76.92\%\\
ALS &1  & 1 & 0& 100.0\%\\
MNFT  &4 & 4 & 0& 100\%\\
LARD & 13 &12  &1 & 92.31\%\\
EPDD & 17 & 17 & 0 &100.0\%\\
\hline
Total & 78 & 71 & 7 &91.03\%\\

\hline

\hline
\end{tabular}
}
\end{table}

\textbf{Recall.} We identified a total of $71$ defects, with a recall rate of $91.03\%$. Our performance was relatively poor on IETI and LARD defects, but using more advanced models could potentially address these shortcomings. Agentable performed poorly in detecting TRE defects with mistakenly identifying three TRE defects. Our analysis of the erroneous IETI detections indicates that a key factor is the LLM’s tendency to make unnecessary judgments. For example, in an TRE defect detection, the LLM sometimes flagged non-essential fault-tolerance aspects missing in the Tool code as IETI defects, such as type checks for intermediate variables that are not required. In practice, such over-detections are unnecessary. Employing a more advanced LLM or designing more effective prompts could help mitigate this issue.

We also analyzed the one instances where Agentable failed to detect LOPE defects. Our analysis revealed that Agentable sometimes struggles with detecting more complex fault-tolerance mechanisms, such as those implemented using if statements. This issue can also be mitigated by designing more effective prompts, such as incorporating few-shot examples. Overall, Agentable demonstrated a relatively high recall rate.

\subsection{RQ3: Characterizing Agent Defects in the Wild}

\begin{figure}[t]
	\centering{\includegraphics[scale=0.29]{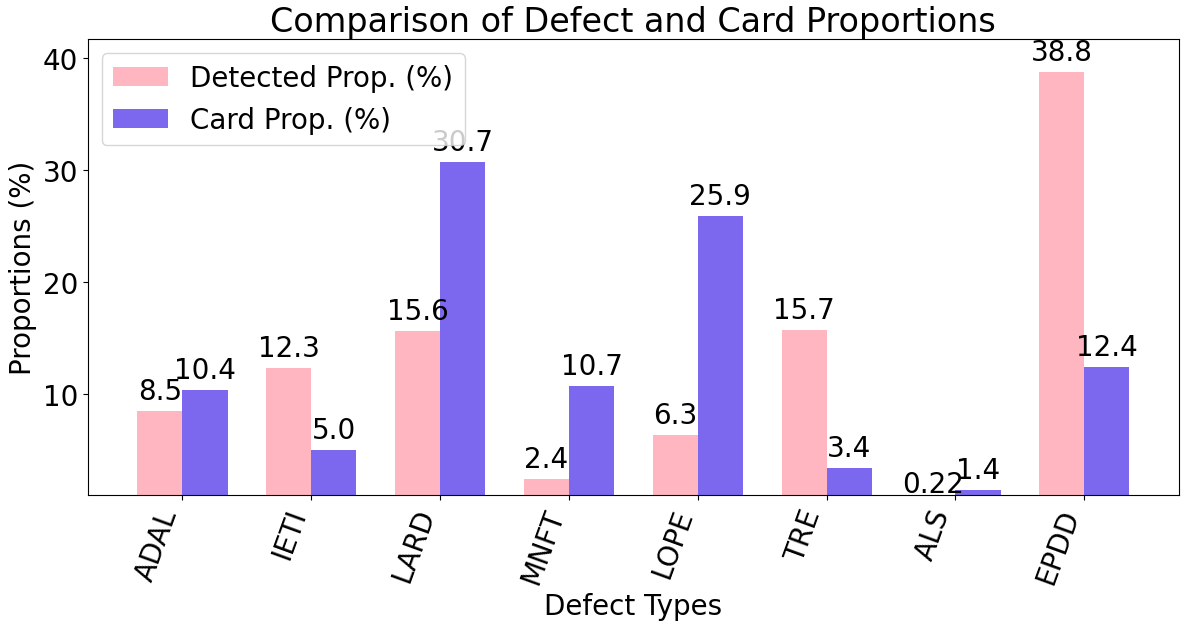}}
	\caption{Defect distribution of Agents.}
	\label{Fig:distri}
\end{figure}

\textbf{Defect Analysis.} While demonstrating the effectiveness of Agentable, our large-scale experiments also provide the first detailed examination of the prevalence and distribution of defects in real-world Agent projects. The first and second columns of Table~\ref{table:rq3} present the number and percentage of defects for each type, while the third and fourth columns show the corresponding data from the empirical analysis in Section~\ref{sec:defect}. As shown in Fig.~\ref{Fig:distri}, among the eight defect types, LARD and EPDD account for the largest proportions in Agentable’s detection results and the empirical data from Section~\ref{sec:defect}. This indicates that defects related to model invocation and package dependencies are the most common defects, both in real-world environments and during the Agent development process. In contrast, ALS defects have the smallest proportion in both datasets, suggesting that this defect type is relatively harder to trigger.

\begin{table}[h]
\centering
\small
\caption{Defect distribution of Agents.}\label{table:rq3}
\scalebox{0.9}{
\begin{tabular}{l||c|c|c|c}
\hline \textbf{Defect} &\textbf{Detected} & \textbf{Detected Prop.(\%)} & \textbf{Card} & \textbf{Card prop.(\%)} \\
\hline 

ADAL & 76 & 8.5 &37 & 10.4\\
IETI  &110 & 12.3 & 18& 5\\
LOPE &56 & 6.3 & 92& 25.9\\
TRE & 140 & 15.7 & 12& 3.4\\
ALS &2  & 0.22 & 5& 1.4\\
MNFT  &21 & 2.4 & 38& 10.7\\
LARD & 139 &15.6  &109 & 30.7\\
EPDD & 345 & 38.8 & 44 &12.4\\
\hline
Total & 889 & N/A & 355 &N/A\\

\hline

\hline
\end{tabular}
}
\end{table}

Additionally, while MNFT and LOPE defects have significant proportion in the empirical data from Section~\ref{sec:defect}, their proportion in Agentable’s detection results is smaller. This suggests that MNFT and LOPE defect are more common during the development process but less evident in well-developed Agent projects. EPDD defects also exhibit notable differences between the two datasets, as Agentable provides warning reports for EPDD defects without necessarily indicating actual defects. ADAL and IETI defects have moderate proportions in both datasets, with a difference of around $9\%$. This implies that while these defects are not highly prevalent, IETI defects are more likely to occur in real-world environments, whereas ADAL defects are more common during development.

\section{Discussion}
\label{sec:discussion}

\subsection{Mitigations for Defects }
\label{subsec:discussion}

During the evaluation, we found that Agent defects are often caused by the relatively loose coupling between the various components of the Agent. Therefore, in addition to presenting Agentable, we also provide possible solutions for each defect type in Table~\ref{table:solution}. These solutions are summarized from the discussion sections of the StackOverflow posts we collected.

\begin{table*}[t]
\centering
\small
\caption{Possible solutions for defects.}\label{table:solution}
\scalebox{0.95}{
\begin{tabular}{l||l}
\hline \textbf{Types} & \textbf{Possible solution} \\
\hline 

ADAL & Design and conduct comprehensive generation capability tests before using an LLM.\\
IETI & Reduce the development of tools with similar functionalities or increase their differentiation.\\
LOPE & Set necessary fault tolerance when invoking an LLM, such as setting handle\_parsing\_error to True in Langchain.\\
TRE & Conduct thorough testing before registering a Tool into the Agent.\\
ALS & Avoid designing overly simple trigger words and avoid using common function return values like None.\\
MNFT & Set necessary fault tolerance for inputs and outputs when the Agent invokes a Tool.\\
LARD & Ensure that the method for invoking the LLM is correct and that no essential parameters are missing.\\
EPDD & Regularly upgrade the Agent framework or test package versions in advance.\\

\hline

\hline
\end{tabular}
}
\end{table*}

\textbf{ADAL.} One potential solution to the AAL defect is to use the best-performing LLM in current benchmarks and thoroughly test its generation and instruction-following capabilities before deployment, based on the selected prompt techniques.

\textbf{IETI.} A potential solution to the ETE defect is to provide detailed descriptions of the roles of different tools in the Agent and ensure as much differentiation as possible between tools with similar functionalities. Additionally, reviewing the development of redundant tools can help avoid the LLM from selecting the wrong tool.

\textbf{LOPE.} A potential solution to the LPE defect is to implement necessary fault tolerance for all inputs and outputs of LLM calls within the Agent. For inputs to the LLM, the type should be checked to ensure they are in a readable format, such as a string. For the LLM’s outputs, checks should be made to ensure they conform to the required format, including type checks.

\textbf{TRE.} A potential solution to the TRE defect is to conduct detailed testing of the tools to ensure they function correctly.

\textbf{ALS.} A potential solution to the ALS defect is to design trigger words with additional care; these words should be uncommon (e.g., function return values) but not overly long or complex, which could affect the quality of the LLM’s generation. Another potential solution to ALS is to adapt all possible return values of the tools and the names of the tools themselves.

\textbf{MNFT.} A potential solution to the MNFT defect is to implement necessary fault tolerance for all inputs and outputs of Tool calls within the Agent. For inputs, the parameter types provided to the Tool should be checked, and for tools that require multiple parameters, the order of the parameters should also be verified. For outputs, additional checks should be performed on the type of tool output.

\textbf{LARD.} A potential solution to the LARD defect is to examine the methods for calling the LLM, especially for Agents that integrate multiple open-source or closed-source LLMs.

\textbf{EPDD.} A potential solution to the EPDC defect is to regularly upgrade the Agent framework and ensure that there are no conflicts in the package versions required by the components of the framework during development.

\subsection{Threats to Validity}
\label{subsec:threats}

\textbf{Internal Threats.} In Section~\ref{sec:defect}, the collection and summarization of defect types, as well as the data collection in Section~\ref{sec:evaluation}, both rely on manual processes. This may introduce inaccuracies. To mitigate this threat, we adopted a hybrid card sorting method that includes a double-checking process, conducted by three authors with over three years of Python programming experience. This approach has been deemed sufficient in prior research.

\textbf{External Threats.} Currently, Agentable only supports Agents built using Python, as Python is the dominant language in the construction of Agent projects and allows better manipulation of models. However, some Agents are still built in other languages, which may limit the applicability of Agentable. One primary solution to address this threat is to adapt Agentable to other programming languages. This would require preparing CPG generators for the target languages and implementing AST parsing methods for the semantic enrichment module, as well as modifying all system prompts. This is primarily an engineering task dependent on the required workload.

\subsection{Limitations}
\label{subsec:limit}

Agentable currently supports detecting standalone Agents but does not support multi-Agent systems. Since the workflows of the two differ significantly, standalone Agents and multi-Agent systems can be considered distinct research subjects. However, our results can still inform the development of multi-Agent systems and detect single-Agent patterns within multi-Agent environments. Another limitation of Agentable is that it can only defect warnings for EPDD defects and cannot fully report them as defects. This is because it is impossible to analyze a function’s dependency on external package versions solely through source code analysis. However, Agentable can provide alerts for this defect, which assists developers in conducting more thorough checks.

\section{Related Work}
\label{sec:related_work}

\subsection{LLM-based Agent Defects}

As LLM Agents are increasingly deployed across various domains of human activity, several studies have explored their security and applicability. Tennant et al. discussed the need for alignment between Agents and human values, as well as testing methodologies~\cite{tennant2024}. Chen et al. explored the possibility of launch-time backdoor attacks against general-purpose and RAG-based LLM Agents through memory poisoning~\cite{chen2024agentpoisonredteamingllmagents}. Tang et al. examined potential issues in scientific work Agents, such as additional memory overhead and regulatory concerns~\cite{tang2024prioritizing}. Li et al. ~\cite{li2024personal} provided a comprehensive review of the development paths and current state of personal LLM Agents, summarizing information on their effectiveness and efficiency.

\textbf{Differences. }The aforementioned research primarily focuses on the capabilities and security flaws of LLM Agents rather than the general programming defects in the workflow of Agents that we are concerned with. The former concentrates on the performance and behavior of LLM Agents, while our focus is on specific defects within the Agent programs to ensure their proper operation.

\subsection{Agents for Detecting Defects in Software}

Several works have explored the application of Agents in software engineering tasks. Yang et al. developed SWE-agent to write code and detect errors~\cite{yang2024sweagent}. Jin et al. conducted a large-scale review of Agents in the field of software engineering, highlighting some of the current drawbacks of Agents~\cite{jin2024llms}. Zhang et al. developed Seeker, an Agent for fixing code vulnerabilities. Rafi et al. enhanced the fault localization capabilities of Agents in code analysis by utilizing self-reflection mechanisms~\cite{rafi2024enhancing}. Another interesting work is Agentless by Xia et al.~\cite{xia2024agentless}, which reduces the context pressure on LLMs by fragmenting code and detects code vulnerabilities using an agentless model. This work also discusses the necessity of Agent models in software engineering tasks.

\textbf{Differences.} The aforementioned research primarily focuses on exploring the use of Agents in software engineering tasks, whereas our work focuses on identifying code defects within the Agents themselves. Our work can help ensure the stability of these studies by providing a more robust foundation for Agent operations.

\section{Conclusion}
\label{sec:conclusion}

In this work, we conducted the first study aimed at understanding and detecting code design defects in LLM-based Agents. Through the analysis of $6,854$ StackOverflow posts, we proposed the first systematic classification of LLM Agent defects, covering eight distinct defect types across key components of Agent workflows. To demonstrate the practical impact of these defects, we introduced Agentable, an LLM-powered static analysis tool designed to detect these defects. By combining Code Property Graphs (CPGs) with LLM-driven analysis, Agentable efficiently identifies complex code patterns and natural language inconsistencies. Our evaluation shows that Agentable achieved an overall precision of $88.79\%$ and a recall rate of $91.03\%$. Additionally, we found that $889$ defect on the real-world Agent projects, highlighting the prevalence of these defects in practice.


\bibliographystyle{IEEEtran}
\bibliography{reference}

\end{document}